%% file: main.tex
\documentclass[conference]{IEEEtran}
\usepackage{enumitem}
\usepackage{multirow}
\usepackage[most]{tcolorbox}
\usepackage{xspace}
\usepackage{ifthen}
\usepackage{amsfonts}
\usepackage{footnote}
\usepackage{url}
\usepackage{colortbl}
\usepackage{tabulary}
\usepackage{etoolbox}

\def\BibTeX{{\rm B\kern-.05em{\sc i\kern-.025em b}\kern-.08em
    T\kern-.1667em\lower.7ex\hbox{E}\kern-.125emX}}
    
\begin{document}

\title{On the Generalizability of Transformer Models\\to Code Completions of Different Lengths}

\input{macros.tex}

\author{
\IEEEauthorblockN{Nathan Cooper\IEEEauthorrefmark{1}, Rosalia Tufano\IEEEauthorrefmark{2}, Gabriele Bavota\IEEEauthorrefmark{2}, Denys Poshyvanyk\IEEEauthorrefmark{3}}
\IEEEauthorblockA{\IEEEauthorrefmark{1}\textit{Stability AI, USA}}
\IEEEauthorblockA{\IEEEauthorrefmark{2}\textit{Universit\`{a} della Svizzera italiana (USI), Switzerland}}
\IEEEauthorblockA{\IEEEauthorrefmark{3}\textit{William \& Mary, USA}}
}

\maketitle

\input{texs/00_abstract.tex}

\begin{IEEEkeywords}
DL4SE, Code Completion
\end{IEEEkeywords}

\input{texs/01_intro.tex}
\input{texs/02_back.tex}
\input{texs/03_relwork.tex}
\input{texs/04_design.tex}

\input{texs/05_results.tex}
\input{texs/06_threats.tex}
\input{texs/07_conclusion.tex}

\section*{Acknowledgment}
This project has received funding from the European Research Council (ERC) under the European Union's Horizon 2020 research and innovation programme (grant agreement No. 851720). The researchers from W\&M have been supported in part by the NSF CCF-2311469 and CNS-2132281. We also acknowledge support from Cisco Systems. Any opinions, findings, and conclusions expressed herein
are the authors' and do not necessarily reflect those of the sponsors.

\bibliography{main}
\bibliographystyle{IEEEtran}

\end{document}

%% file: macros.tex
%%%%%%%%%%%%%% @danaderp
%extra commands

\definecolor{lightgrey}{HTML}{d3d3d3}
\newtcolorbox{resultbox}{colback=lightgrey, arc=0.5mm, top=2mm, bottom=2mm, left=2mm, right=2mm}

\newboolean{showcomments}

\setboolean{showcomments}{true}

\ifthenelse{\boolean{showcomments}}
  {\newcommand{\nb}[2]{
    \fbox{\bfseries\sffamily\scriptsize#1}
    {\sf\small$\blacktriangleright$\textit{#2}$\blacktriangleleft$}
  }
  \newcommand{\cvsversion}{\emph{\scriptsize$-$Id: macro.tex,v 1.9 2005/12/09 22:38:33 giulio Exp $}}
  }
  {\newcommand{\nb}[2]{}
  \newcommand{\cvsversion}{}
  }

\newcommand\myparagraph[1]{\noindent\underline{\bf {#1}:}}
\newcommand\myparagraphnew[1]{\noindent{\bf {#1}:}}
\newcommand\emparagraph[1]{\noindent {\em {#1}:}}
\newcommand\budget[1]{{\color{red}\myparagraph{Budget}{#1} pages}}
\newcommand\rosalia[1]{{\color{ForestGreen} \nb{ROSALIA}{#1}}}
\newcommand\denys[1]{{\color{blue} \nb{DENYS}{#1}}}
\newcommand\gabriele[1]{{\color{brown} \nb{GABRIELE}{#1}}}
\newcommand\nathan[1]{{\color{violet} \nb{NATHAN}{#1}}}
\newcommand\TODO[1]{{\color{red} \nb{TODO}{#1}}}

\newcommand{\cancel}[1]{{\leavevmode\color{RubineRed}{\sout{\xspace#1}}}}
\newcommand{\edit}[2]{{\leavevmode\color{RubineRed}{\sout{#1}}}{\color{blue}{\xspace#2}}}
\newcommand{\rewrite}[2]{{\leavevmode\color{RubineRed}{\sout{#1}}}{\color{Green}{\arrow\xspace#2}}}

\newcommand{\add}[1]{{\leavevmode\color{black}{\xspace#1}}}
\newcommand{\addnew}[1]{{\leavevmode\color{Green}{\xspace#1}}}
\newcommand{\remove}[1]{{\leavevmode\color{red}{\xspace#1}}}

\newcommand\finding[1]{\vspace{0.25em}\noindent\textsf{\bf Finding {#1}.}}
\newcommand\fnumber[1]{{$\mathcal{F}_{#1}$}}
\newcommand\operator[2]{{\bf OP$_{#1}$: {\em {#2}} -- }}
\newcommand\opnumber[1]{{{\bf OP}$_{#1}$}}
\newcommand{\arrow}{{$\rightarrow$}\xspace}
\newcommand\inline[1]{{\lstinline{#1}}}

\newcommand{\boxme}[1]{{
\begin{tcolorbox}[enhanced,skin=enhancedmiddle,borderline={1mm}{0mm}{MidnightBlue}]
    \textbf{Take Aways: } #1 \end{tcolorbox} 
}}

\newcommand\fix[1]{{\color{blue} \nb{FIX THIS}{#1}}}
\newcommand\blue[1]{{\color{blue}{#1}}}
\newcommand{\here}{{\color{blue} \nb{***}{CONTINUE HERE}}}

\newcommand{\REFF}{\textcolor{red}{\textbf{[REF]}}}

\newcommand{\REF}{{\color{red} \textbf{[REFS]}}\xspace}
\newcommand{\xy}{{\color{red} \textbf{XY}}\xspace}
\newcommand\tops[1]{{\color{blue}{#1}}}
\newcommand\alert[1]{{\color{red}{#1}}}

\newcommand{\target}{\textit{target tool}\xspace}
\newcommand{\targets}{\textit{target tools}\xspace}
\newcommand{\behavior}{\textit{target behavior}\xspace}
\newcommand{\ie}{\textit{i.e.,}\xspace}
\newcommand{\eg}{\textit{e.g.,}\xspace}
\newcommand{\etc}{\textit{etc.}\xspace}
\newcommand{\etal}{et al.\xspace}
\newcommand{\etals}{et al.'s\xspace}
\newcommand{\aka}{\textit{a.k.a.}\xspace}	

%%%%%

\newcommand{\secref}[1]{Sec.~\ref{#1}\xspace}
\newcommand{\figref}[1]{Fig.~\ref{#1}\xspace}
\newcommand{\tabref}[1]{Table~\ref{#1}\xspace}
\newcommand{\Phase}{{\sc Phase}\xspace}
\newcommand{\Phases}{{\sc Phase's~}\xspace}

\newcommand{\emphquote}[1]{{\emph{`#1'}}\xspace}
\newcommand{\emphdblquote}[1]{{\emph{``#1''}}\xspace}

\newcommand{\emphbrack}[1]{\emph{[#1]}\xspace}
			
\newcommand{\subj}{\emphbrack{subject}}
\newcommand{\act}{\emphbrack{action}}
\newcommand{\obj}{\emphbrack{object}}
\newcommand{\prep}{\emphbrack{preposition}}
\newcommand{\objtwo}{\emphbrack{object2}}

\newcommand*\ciclednum[1]{\raisebox{.5pt}{\textcircled{\raisebox{-.9pt}
{#1}}}}

%----------------------

%% file: texs/00_abstract.tex
% !TEX root = ../main.tex
\begin{abstract}
The programming landscape is nowadays being reshaped by the advent of Large Language Models (LLMs) able to automate code-related tasks related to code implementation (\eg code completion) and comprehension (\eg code summarization). Such a paradigm shift comes with a number of implications related to how software will be written, maintained, and evolved. Also, these LLMs are extremely expensive to train, posing questions on their sustainability over time. Given their training cost, their ability to generalize, namely their ability to work on task instances different from those on which they have been trained, is an aspect worth being investigated. 
Previous work already showed that transformer models can successfully support code completion in a cross-project setting. However, it is unclear whether LLM are able to generalize to inputs having lengths not seen during training.  For example, it is known that training a model on short instances allows to substantially reduce the training cost. However, the extent to which such a model would provide good performance on sequences having lengths not seen during training is not known. Many recent works in Natural Language Processing (NLP) tackled this problem in the context of decoder-only LLMs, \ie xPOS and ALiBi. To assess if these solutions extend to encoder-decoder LLMs usually adopted in the code-related tasks, we present a large empirical study evaluating this generalization property of these and other encoding schemes proposed in the literature, namely Sinusoidal, xPOS, ALiBi, and T5. We found that none of these solutions successfully generalize to unseen lengths and that the only safe solution is to ensure the representativeness in the training set of all lengths likely to be encountered at inference time. %We discuss different implications of our findings and the trade-offs researchers and developers of LLMs for code completion can make. %We release our code, datasets, and models to foster research in this area.
% the length extrapolation ability of these solutions (\ie trained on small code instances}

% there are two important generalizability aspects that remain unexplored. First, the \emph{complexity} of the code completion task: Can a model trained on simple code completion problems (\eg predicting the next token to type) properly work in more complex scenarios (\eg predicting entire code blocks) and \emph{vice versa}? Second, the \emph{length} of the code instances used for training: Can a model trained on code completion problems within short code instances be successfully applied to predict code tokens in long code instances and \emph{vice versa}? To answer these questions, we study the generalizability in terms of \emph{complexity} and \emph{length} of ALiBi and T5, two types of positional encoding schemes for the Transformer architecture that, in the Natural Language Processing field, showed the ability to train on short sequences, thereby being more efficient to train, and generalize to long sequences without a large performance degradation. Our results show that, while T5 better supports code completion, both schemes lack the ability to generalize when trained and tested on instances having different \emph{complexity} and \emph{length}. We release our code, datasets, and models to foster research in this area.
\end{abstract}

%% file: texs/01_intro.tex
% !TEX root = ../main.tex
\section{Introduction}\label{sec:intro}

Large Language Models (LLMs) for code have achieved state-of-the-art results across a variety of software engineering (SE) tasks such as code completion \cite{WhiteMSR2005, ciniselli2021empirical}, code reviews \cite {TufanoICSE22}, code summarization \cite{leclair-mcmillan-2019-recommendations}, program repair \cite{TufanoTOSEM2019, Chen:2019}, testing \cite{WatsonAsserts,TufanoMutation}, and others \cite{watson2022systematic,MastropaoloT5,TufanoMeaningfulChanges}. The Transformers \cite{vaswani2017attention} have been at the center of these improvements due to its attention mechanism and ability to be highly parallelized, allowing for more efficient training compared to previous models such as Recurrent Neural Networks (RNN) \cite{rumelhart1985learning}.

When applied to code completion, Transformers take as input an incomplete code component and try to predict the missing code tokens (\eg \cite{ciniselli2021empirical, chen2021evaluating, austin2021program, fried2022incoder}). As already happened in the field of Natural Language Processing (NLP) \cite{press2020shortformer}, there has been a recent push in increasing the maximum length of the sequences (incomplete code components) on which Transformers are trained and tested. 

This is due to the fact that longer sequences (i) allow to provide the model with additional contextual information which can help with improving the prediction performance; and (ii) can help in simulating more variegate code completion scenarios. This, however, has a substantial cost to pay in terms of training time \cite{press2022alibi}.

It comes without surprise that efforts have been made in the NLP literature to address this issue (\eg \cite{dai2019transformer,raffel2020exploring,press2022alibi,sun2022length}): The most recent work targets the generalization of Transformers to longer sequences than those they have been originally trained for \cite{press2022alibi,sun2022length}. This allows to efficiently train a model on short sequences and, then, perform the inference on longer sequences without a significant performance degradation. This is, for instance, the goal of the ALiBi (Attention with Linear Biases) attention mechanism for Transformers \cite{press2022alibi}, which has been successfully used in NLP.

%Some approaches have attempted to overcome this issue such as Transformer-XL \cite{dai2019transformer}, which added a recurrent mechanism to the standard Transformer, or using relative positional embeddings over absolute positional embeddings have been shown to allow for generalization to larger context windows than those seen during training \cite{press2022alibi}. The most notable being the T5 relative position encoding \cite{raffel2020exploring}. However, these add a significant performance hit, both in terms of memory and speed, in training and inference that counteracts any gains of being able to train on short sequences and test on long sequences. Therefore, Press \etal \cite{press2022alibi} introduced a novel bias to the Transformer attention mechanism called Attention with Linear Biases (ALiBi) that allows for efficient training on short sequences and generalizing to long sequences during inference without a significant performance degradation.

If solutions such as ALiBi properly work on source code as well, they could substantially help reduce the training cost of code completion models such as the popular GitHub Copilot \cite{github}. This is the focus of our work. We aim to investigate the extent to which solutions proposed in the NLP literature can support the generalization of Transformers on source code. We focus on three state-of-the-art solutions and one baseline. \add{The first (baseline), Sinusoidal \cite{vaswani2017attention}, uses Absolution Positional Encodings (APEs) by defining sine and cosine functions to generate positional embeddings that the authors original hypothesized would help the model to generalize. The second, xPOS \cite{su2021roformer,sun2022length}, is a hybrid between APEs and Relative Positional Encodings (RPEs) and applies rotations to the sine and cosine positional embedding to incorporate relative position information along with a attention resolution metric to improve generalization. The third, ALiBi \cite{press2022alibi}, offers a simple solution of modifying the attention mechanism to weight positions far away as less important than ones closer. The last, T5 \cite{raffel2020exploring}, similarly modifies the attention mechanism by adding a learned bias term that influences the attention given to a token.}

% The first, is the already introduced ALiBi positional encoding scheme \cite{press2022alibi}, while the second one is the Text-To-Text-Transfer-Transformer (T5) model proposed by Raffel \etal \cite{raffel2020exploring} which uses a different relative positional encoding scheme for generalization to larger context windows than those seen during training.

% When dealing with code completion tasks, there are two aspects of generalizability we are interested in investigating. The first is the one just discussed, namely the \emph{length} of the instances on which the model is trained and tested.
We want to assess whether models trained on sequences of a specific length are able to generalize, \ie not incur a significant performance degradation, on sequences being longer (or shorter) than the training ones. To accomplish this, we built four datasets (\emph{short}, \emph{medium}, \emph{long}, and \emph{mix}) featuring incomplete Python and Java functions having different lengths. 

\eject

Then, we train 32 Transformers, namely four models (Sinusoidal, xPOS, ALiBi, and T5) for each of the four datasets and two programming languages. The performance of the 32 models has been evaluated on a series of test sets of previously unseen Python and Java functions of various different lengths, studying the generalization of their predictions. For example, we verified if the models trained on \emph{short} datasets are able to work on instances in the test set having a length inline with the examples in the \emph{long} dataset.

Overall, we make the following contributions:
\begin{enumerate}
    \item \add{A systematic benchmark for evaluating the generalization of LLMs for code completion of different lengths across two programming languages;}
    \item \add{An empirical study on whether current generalization approaches extend to encoder-decoder architectures for the task of code completion;}
    \item \add{A set of results and implications that can be leveraged by researchers and developers of these models for navigating trade-offs.}
\end{enumerate}

% Our findings can be summarized as follows:
% \begin{enumerate}
%     \item Relative performs significantly better than all of the other model types in the task of code completion, independently from the code instance \emph{length};
%     \item None of the models studied have a strong ability to generalize to unseen \emph{lengths};
%     % \item Training on a mixture of \emph{complexities} ensures good performance on instances of any complexity seen during training, with performance on par or even slightly better than what can be achieved with a specialized model (\ie a model only trained to work on specific complexities);
%     \item \add{A large empirical study on whether the supposed approaches extend to encoder-decoder architectures}
%     \item \add{a set of results that can be leveraged by developers of these models as a design matrix that gives suggestions for which model to use given requirements}
%     \item However, surprisingly, training on a mixture of \emph{lengths} \textbf{does} harm performance significantly as compared to models specialized to work on specific instance lengths.
% \end{enumerate}

% Our findings suggest that the selection of the complexity and length of the training instance is a crucial choice with major implications on the performance of the trained models in different complexity/length scenarios.

%% file: texs/02_back.tex
% !TEX root = ../main.tex
\section{Background}\label{sec:back}

\begin{figure}[t]
    \begin{center}
		\includegraphics[width=0.73\linewidth]{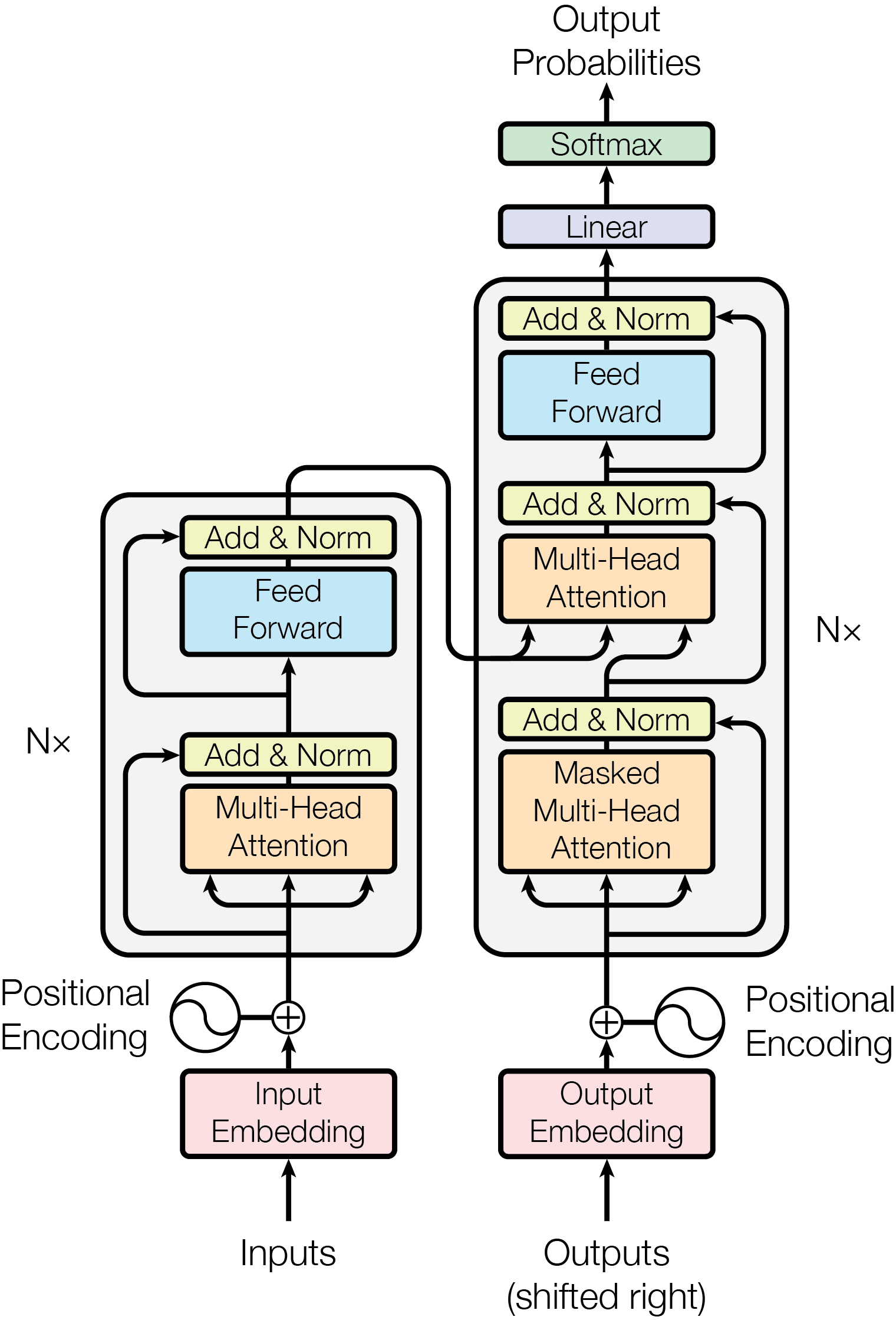}
        \caption{Sequence to Sequence Transformer Overview from the original paper \cite{vaswani2017attention}. The left part is the encoder and the right part is the decoder.}
        \label{fig:transformer}
    \end{center}
\end{figure}

\begin{figure}[t]
    \begin{center}
		\includegraphics[width=0.9\linewidth]{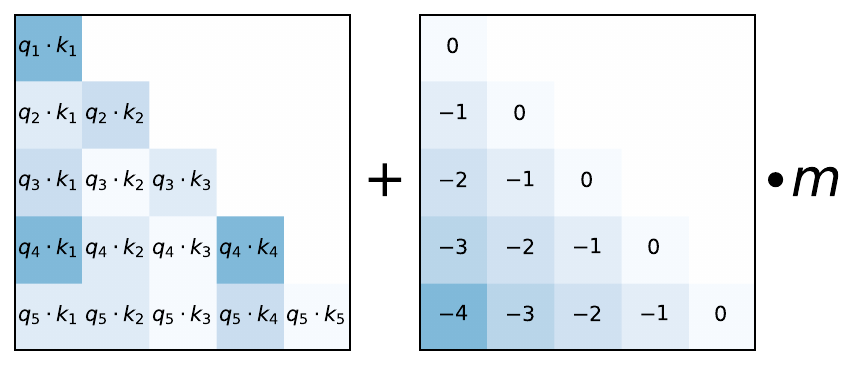}
        \caption{ALiBi Overview from the original paper \cite{press2022alibi}.}
        \label{fig:alibi}
    \end{center}
\end{figure}

We introduce some mathematical background to understand the specifics of the position encoding schemes, namely Sinusoidal, xPOS, ALiBi, and Relative, and how they apply to our study.
Our paper focuses on the Sequence to Sequence Transformer \cite{vaswani2017attention} (see Fig. \ref{fig:transformer}). It takes as input a sequence of tokens $C'={x_1,x_2,...,x_n}$ (in our case, the code to be completed), and outputs a target sequence, $M$, that is similarly decomposed into tokens $M={y_1,y_2,...,y_m}$. 

$M$ represents in our context the missing piece of code to be predicted. Thus, the combination $C = C' + M$ equals the complete code snippet. Note that the $+$ operator does not imply an append, but rather a combination regardless of where the missing part $M$ is placed in $C'$. The decomposition of both $C'$ and $M$ is done through tokenization using a trained Byte Pair Encoding tokenizer \cite{sennrich2015neural} such that $C' \text{ or } M \rightarrow s_0,s_1,...s_T$, where $s_i \in \mathbb{V}$ and $\mathbb{V}$ is the vocabulary of the tokenizer. We use the same vocabulary for both $C'$ and $M$. These tokens are passed through an Embedding layer, which is shared across the encoder and decoder, to get the token's vector representation, which will be updated progressive through various Attention and Feed Forward layers, we will explain further.

The output $M$ is done in an autoregressive manner, where the output of one time step from the model is fed back into the decoder portion of the Transformer for generating the probability distribution of the next token. Formally, the probability distribution of token $y_i$ is conditioned on the output of the encoder $Z$ and the $y_{<i}$ previously generated tokens: $p(M)=\prod^m_i{p(y_i|y_{<i}, Z)}$.

% We do not discuss the details of the base Transformer architecture which are out the focus on the modifications of ALiBi and T5 that have been shown to help with generalization in natural language processing. We kindly refer the interested reader to the original Transformer paper \cite{vaswani2017attention} for further details on the Transformer architecture.

The Transformer architecture itself has no way of modeling sequential information. Therefore, sequential information is injected either at the bottom of the network (Fig. \ref{fig:transformer}) or at each Transformer block of the network depending on the positional scheme. For our four positional schemes, only one injects the positional information at the bottom of the network, namely sinusoidal, the rest of the schemes inject the information at each Transformer block.

% The two schemes we investigate both inject positional information at each Transformer block by modifying the computation of the self-attention mechanism. However, we will first explain the original Transformer's way of computing the attention. 

The Transformer block's composition depends on if it is part of the encoder or decoder. In the encoder portion, the
% The two schemes we investigate both inject positional information at each Transformer block by modifying the computation of the self-attention mechanism. In the original Transformer,
attention is computed by transforming a given sequence, $s=(s_1,s_2,...,s_n)$ where $s_i \in \mathbb{R}^{d_s}$ into a new sequence of the same size, $z=(z_1,z_2,...,z_n)$ where $z_i \in \mathbb{R}^{d_z}$ via a weighted sum where the weight can be intuitively thought as the amount of attention to pay to the value of $s_j$ in the sequence. $\mathbb{R}^{d_s}$ represents the embedding space of the Embedding layer that transforms the discrete token into a continuous vector of size $d_s$ and similarly $\mathbb{R}^{d_z}$ represents the vector space that is composed of $d_z$ dimensions where $d_z$ can be the same or a different size than $d_s$. The weighted sum can be represented as follows:

\begin{equation}
    z_i=\sum^n_{j=1}a_{ij}(s_jW^V)
\end{equation}

With $W^V$ being a learned value weight matrix and $a_{ij}$ being calculated with the following softmax formula:

\begin{equation}
    a_{ij}=\frac{\text{exp }e_{ij}}{\sum^n_{k=1}\text{exp }e_{ik}}
\end{equation}

And $e_{ij}$ being calculated by taking the corresponding $s_i$ and $s_j$ tokens and multiplying them by a learned query and key weight matrix, $W^Q$ and $W^K$, to get the corresponding query and key vectors. These vectors are then put through a compatibility function, namely the scaled dot product:

\begin{equation}
    e_{ij} = \frac{(s_i W^Q)(s_j W^K)^T}{\sqrt{d_k}}
\end{equation}

As it can be seen, no positional information is present in these operations. So, the original Transformer injected positional information into the beginning of the network by adding a position vector to the token vector to encode this information for the rest of the network. Let us now discuss each positional encoding scheme.\smallskip

\add{\textbf{Sinusoidal:} Sinusoidal is the original scheme proposed in the Transformer paper \cite{vaswani2017attention}. The information is added directly to the token embeddings at the beginning of the network. Concretely, the same dimension of the token embeddings is used for the position embedding and the sinusoidals switch between sine and cosine:}

\begin{equation}
    PE_{pos, 2i} = sin(pos/10,000^{2_i / d_{model}})
\end{equation}

\begin{equation}
    PE_{pos, 2i + 1} = cos(pos/10,000^{2_i / d_{model}})
\end{equation}

\add{Where $pos$ refers to the position in the sequence and $i$ refers to the specific dimension of the position embedding. The authors chose this scheme as they believed it would allow for the model to learn to use relative positions. However, this approach is generally considered an Absolute Positional Encoding (APE) scheme.}\smallskip

\add{\textbf{xPOS:} Sun \etal \cite{sun2022length} extended work from Su \etal \cite{su2021roformer} which  made the observation that the dot product between queries and keys are where information is shared between different tokens. Therefore, position information can also be added of the relative position between the different tokens. Specifically, Su \etal \cite{su2021roformer} wanted to find an operation that satisfies the following:}

\begin{equation}
    (f_q(x_m, m), f_k(x_n, n)) = g(x_m, x_n, m - n)
\end{equation}

\add{Where functions $f_q$ and $f_k$ add this relative information to the token embeddings $x_m$ and $x_n$. respectively. To accomplish this, the authors introduced Rotary, which uses rotations of the token embeddings based on their position so that the relative position of the tokens are preserved through this dot product. Namely, they define the function $g$ that satisfies this to be}

\begin{equation}
    g(x_m, x_n, m - n) = Re( (W_q \cdot x_m) (W_k \cdot x_n) * e^{i(m - n) \theta } )
\end{equation}

where the function $Re$ takes only the real part of the complex number. This is specifically for the 2D case, however, this is generalizable to any dimension that the token embeddings belong to. 

For a complete discussion of the equations and their generalization, we refer readers to the original paper \cite{su2021roformer}.

\add{Sun \etal \cite{sun2022length} extended this approach to have similar extrapolation abilities of ALiBi, discussed below, while still having better performance. To accomplish this, the authors introduced the idea of \textit{attention resolution} where a model's attention should monotonically decrease as the pair wise distance between tokens increases, similar to ALiBi.

To integrate this into the rotation matrix, they apply an exponential decay that adds this property. They show that this gives a good trade-off between the Rotary performance and ALiBi's ability to extrapolate to longer than seen during training sequences. For our experiments with xPOS, we use the original implementation from Su \etal \cite{su2021roformer} for the encoder and the extension, xPOS, by Sun \etal \cite{sun2022length} for the decoder since the extension is unable to be applied directly to an encoder.}\smallskip

\textbf{ALiBi:} In ALiBi, the positional information is injected by modifying the equation above by adding a static bias:

\begin{equation}
    e_{ij} = \frac{(s_i W^Q)(s_j W^K)^T}{\sqrt{d_k}}+m^h(j-i)
\end{equation}

Where $m$ is a head-specific scalar that is selected before training. We use the same geometric series for initializing these $m$ values per head as in the original ALiBi paper, namely starting from $2^\frac{-8}{n_{heads}}$ and using the same value as the ratio. Intuitively, this static bias penalizes query and key vectors that are far away from each other. Figure \ref{fig:alibi} visualizes this process. Specifically, it shows how queries and keys corresponding to the same token do not receive any reduction whereas mismatched queries and keys receive a reduction proportional to their relative distance.

This process was originally designed for decoder-only Transformer models. However, since we use an encoder-decoder Transformer model, we additionally use the bidirectional version for the encoder portion as outlined in a post by the original author \footnote{\url{https://github.com/ofirpress/attention_with_linear_biases/issues/5}}.\smallskip

\textbf{T5:} Similar to ALiBi, T5 introduces a bias inside the softmax equation that is based on distance. Specifically, this bias is a learned scalar that is added to the query and key dot product $(s_i W^Q)(s_j W^K)^T+b^h_{ij}$ where each attention head, $h$, has a different learned bias. T5 introduces this idea of buckets, which are the different learned biases, where the different $ij$ pairs logarithmically map up to a relative position of 128 beyond which the same $ij$ pairs are mapped to the same bucket. After each attention operation in the encoder Transformer block follows a normalized residual connection, a simple Feed Forward Multi-Layer Perceptron, and another normalized residual connection. A similar process happens in the decoder block.

However, there is an additional attention mechanism that happens after the normal self-attention is applied to the transformed output token embeddings, which considers the encoder's representation of the input, $Z$. This is known as cross-attention and follows the same process as self-attention except the keys and values are constructed from the representation $Z$ while the query is built from the output representation. Intuitively, this can be thought of as the output tokens requesting specific information from the input tokens.

%% file: texs/03_relwork.tex
% !TEX root = ../main.tex

\section{Related Work}\label{sec:relwork}

We discuss the literature related to (i) code completion, (ii) techniques aimed at improving the generalizability of Transformers, and (iii) NLP studies aimed at investigating the extent to which Transformers can generalize to instances different from those seen during training.
% \denys{@Nathan: please update the related work with any recent SE/NLP papers}
\subsection{Code Completion} Code completion has been studied extensively for several years in SE \cite{hindle2016naturalness, tu2014localness, hellendoorn2019code, raychev2014code, bruch2009learning, jin2018hidden, izadi2022codefill, aye2021learning, hellendoorn2017deep, hou2010towards, nguyen2019combining, robbes2008program, robbes2010improving, svyatkovskiy2021fast, svyatkovskiy2019pythia, yang2017language, chen2021evaluating, ciniselli2021empirical, austin2021program, fried2022incoder, nijkamp2022codegen, allal2023santacoder}. It has seen many iterations going from techniques able to generate simple predictions such as method calls \cite{mandelin2005jungloid} to recent DL models able to predict multiple code statements \cite{chen2021evaluating, ciniselli2021empirical, fried2022incoder, nijkamp2023codegen, allal2023santacoder}.

Our work can be thought of as a research thrust continuation to that one of Ciniselli \etal \cite{ciniselli2021empirical}. In their work, they explored the applicability of Transformer models for the task of code completion, especially as the number of tokens to complete grows. They found a T5 architecture to perform fairly well on this task, reporting however a performance degradation as the number of tokens to predict grew. While Ciniselli \etal \cite{ciniselli2021empirical} study how a T5 model trained on a specific dataset can work with prediction tasks of different complexity, we study how models trained on different datasets featuring instances characterized by different lengths generalize to unseen lengths. 

Hendrycks \etal \cite{hendrycks2021measuring} and Chen \etal \cite{chen2021evaluating} proposed a systematic evaluation for code generation tools using functional-correctness. However, their focus was on generating complete programs rather than completing existing code. 

Fried \etal \cite{fried2022incoder} investigated a novel infilling pretraining scheme for decoder-only Transformer architectures that allow them to use bidirectional context to complete code. They found this infilling scheme allows for models that achieve a higher code completion rate than left context only models for single and multi-line code completions. Our study is complementary to these since we focus on encoder-decoder models and on the generalization to code completion length, rather than general performance.

\subsection{Methods to Improve Length Generalization} There has been a plethora of literature on different techniques to improve generalization of inference length of Transformer models. Neishi and Yoshinaga \cite{neishi2019relation} demonstrated that replacing the positional encoding scheme in Transformers with a Recurrent Neural Network (RNN) improves machine translation performance on sentences longer than those seen during training. In a similar vein, Dai \etal \cite{dai2019transformer} take inspiration from RNNs by adding a segment-level recurrence mechanism to improve performance on long sequences. Dubois \etal \cite{dubois2020location} introduced a separate location and content based attention to improve generalization to longer sequences than those seen during training. Newman \etal \cite{newman2020eos} showed that training models to predict the end of sequence (EOS) token significantly degraded a model's ability to generalize to longer sequences than those seen during training. 

Specifically, they found that the hidden states of models trained on predicting EOS tokens lead to stratification of the hidden state manifold by length and get clustered into \textit{length attractors}, which are areas where the EOS token is given the highest-probability prediction. This causes a failure to generalize to longer sequences that are not present when omitting the prediction of the EOS token. Lastly, several works \cite{kiyono2021shape, likhomanenko2021cape, press2022alibi, su2021roformer, sun2022length} introduced various modifications to the positional encoding schemes of Transformers to improve generalization to longer sequences not seen during training. Among those, we considered in our study the four described in Section \ref{sec:back} due their good representativeness of the different types of encoding schemas, namely Absolute Positional Encoding (APE) and Relative Positional Encoding (RPE). In addition, we considered the T5 model since Press \etal \cite{press2022alibi} also showed it had an ability to slightly generalize to longer sequence lengths than it had seen during training.

\subsection{Evaluations of Length Generalization} The most similar work to ours (from the NLP field) is Rosendahl \etal \cite{rosendahl2019analysis}'s study on analyzing a variety of positional encoding schemes and their ability to generalize to longer machine translation sentences than those seen during training. Similar to other works \cite{neishi2019relation, press2022alibi}, they found that relative positional encodings are superior to absolute positional encodings for generalizing to longer sequences. Lake and Baroni \cite{lake2018generalization} and Hupkes \etal \cite{hupkes2020compositionality}'s work focused on measuring the composability of language models. One type of composition was specific to generalization of sequence length and they both found current language models to be unable to perform well on such tasks. %It may be that our task of code completion requires generalization skills more akin to composability rather than to standard language modeling and machine translation. 

%This may account for our results being in opposition to these works that relative position encoding allows for generalizing to longer sequences in those tasks. %More research will need to be done in this area.

% Interestingly, we did not find our task of code completion to benefit to the type of generalizability of relative positional encodings that were demonstrated in machine translation and language modeling. We suspect code completion requires a different type of generalizability akin to the systematic and compositional length generalizability discussed and measured in Lake and Baroni \cite{lake2018generalization}'s and Hupkes \etal \cite{hupkes2020compositionality}'s work. It is an open question the composibility of code, but one potential reason for our findings in opposition to other work is the requirement of the language model to have learned skills composibility to solve code completion lengths outside the ones it saw during training. More research will need to be done in this area.

% In Lake and Baroni, they designed the benchmark SCAN to evaluate the composability skills of language models, which included generalizing to longer sequences. Only work by Newman \etal \cite{newman2020eos} evaluated on this benchmark, as other approaches only 

% Lake and Baroni \cite{lake2018generalization}

% There has been a lot of literature attempting to come up with 

%% file: texs/04_design.tex
% !TEX root = ../main.tex
\section{Evaluation Design}\label{sec:design}

The \emph{goal} of our study is to determine whether popular positional encoding schemes for length generalization work for the task of code completion. Namely, we seek to answer the following question:

\begin{enumerate}[label=\textbf{RQ$_{\arabic*}$:}, ref=\textbf{RQ$_{\arabic*}$}, wide, labelindent=5pt]\setlength{\itemsep}{0.2em}
      \item \label{rq:extrapolate}\textit{To what extent can different positional encoding schemes generalize to different code lengths for the task of code completion?}
\end{enumerate}

In the context of code completion this means studying whether models trained on completing code sequences having a specific length generalize when used to complete shorter/longer sequences. 

This is analogous to whether models are able to utilize different amounts of information, in terms of the input tokens, as compared to what they have seen during training. Naturally, shorter instances require shorter training time. Yet, it is unclear if a model trained on short code completions can generalize to also work on longer ones and \textit{vice versa}.
While answering RQ$_1$ we also check whether our findings generalize to multiple programming languages. In particular, we contextualize our study to Python and Java, as languages  often adopted in code completion studies (see \eg \cite{ciniselli2021empirical,svyatkovskiy2021fast}). Also, we consider two different code completion tasks recently used in the code completion study by Ciniselli \etal \cite{ciniselli2021empirical}: \emph{statement-level} and \emph{block-level} completion. 

The former is the classic code completion task in which the last $n$ tokens of a single code statement are masked, with the model in charge of predicting them. The latter, instead, possibly extends the completion to multiple statements, masking the last $n$ tokens in a block of code (\eg the body of an \texttt{if} statement) and asking the model to guess them.

In the following we detail the procedure used to (i) collect the training/testing datasets employed in our study (Section \ref{subsec:data}), and (ii) perform the required data collection and analysis (Section \ref{subsec:experiments}).

% are able to generalize to the task of code completion. 

% We wish to determine the ability of popular Transformer models to generalize to code completion \emph{complexities} and \emph{lengths} different than those seen during training. To this end, we sought to answer the following research question (RQ):

% % \textbf{$RQ_1$:} \textit{What abilities do current Transformers have to generalize to code completions of different length than those seen during training?} % For this RQ, we are interested in 

% \begin{enumerate}[label=\textbf{RQ$_{\arabic*}$:}, ref=\textbf{RQ$_{\arabic*}$}, wide, labelindent=5pt]\setlength{\itemsep}{0.2em}
%       \item \label{rq:generalize}\textit{To what extent can Transformers generalize to code completions having different characteristics than those seen during training?}
% \end{enumerate}

% We focus on two specific characteristics, namely the \emph{complexity} and the \emph{length} of code completions. Complexity is approximated as the quantity of code tokens that the model is asked to predict. Intuitively, predicting entire code statements is easier than predicting the next single token the developer is likely to write. The length of the code completion instances is, similar to what is done in NLP, simply the length of the instances on which the model is required to perform the prediction. Naturally, shorter instances require shorter training time. Yet, it is unclear if a Transformer trained on short code completions can then generalize to also work on longer ones.

\subsection{Dataset Construction}\label{subsec:data}

To build the datasets needed for our study, we mined data from GitHub open source projects. In particular we: (i) used the GitHub search tool by Dabi\'c \etal \cite{Dabic:msr2021data} to identify all Python and Java GitHub projects having at least 100 commits, 10 contributors, 10 stars and 10 issues (to exclude toy projects); (ii) sorted the projects by number of stars; (iii) cloned the top 3k and extracted from each of them the functions in the master branch (to only rely on functions likely to be syntactically correct); (iv) removed all functions containing non-ascii characters (to avoid problems when reading data); (vi) removed all duplicates (to avoid leaking of information between the training, validation, and test sets we created out of this dataset); (vii) removed from the dataset all instances consisting of more than 1,024 tokens. The latter is a procedure usually adopted in applications of DL4SE (\eg \cite{leclair-mcmillan-2019-recommendations,Tufano:testGen,Chen:2019}). Indeed, too long instances make the training of DL models too expensive, also motivating our investigation into the length generalizability. 

Such a process resulted in the collection of $\sim$4M Python functions and $\sim$4.5M Java functions which we further processed to create datasets aimed at answering our RQ.

\subsubsection{Java dataset: statement-level code completion task}
The Java dataset will cover the statement-level code completion task. The goal is to build three datasets featuring instances (\ie functions)  having different lengths, namely \emph{short}, \emph{medium}, and \emph{long} instances. Basically, with ``length'' we refer to the number of input tokens provided to the model. In all three datasets we keep constant the complexity of the prediction to generate (\ie the number of masked tokens). Indeed, only in this way we can ``isolate'' the impact on performance of changing the input length. 

Since the code completion task, which we are focusing on, requires the masking of code statements, we start by removing for the set of collected Java functions all of those that do not contain any statements. We also removed all Javadoc comments, since we are focusing on completion tasks within the body of a function. We then compute the number of Java tokens within each of the remaining 3,833,445 functions: this results in a distribution of functions' \emph{length}. 

We then compute a second distribution representing the number of Java tokens within the code statements in the subject functions, observing a median of 11 tokens per statement. The idea is to mask in each instance of the three datasets we will create (\ie \emph{short}, \emph{medium}, \emph{long}) the exact same number of tokens (11). 

As previously mentioned, this is done to keep constant the ``complexity'' of the completion task and better isolate the impact of the sequence length on the observed performance.

Given such a constraint, we remove all the functions not containing any valid statements to mask, \ie a statements consisting of at least 11 Java tokens. We sort the remaining 1,855,578 functions by their length and split them into three sets of the same size, obtaining: (i) a first set of functions with lengths ranging from 6 to 96 (\emph{short} dataset); (ii) a second set of functions with lengths ranging from 97 to 180 (\emph{medium}); and (iii) a third set of functions with lengths ranging from 181 to 1024 (\emph{long}).

For each function $F$ in each of the three datasets, we created $n$ training instances, where $n$ is the number of valid statements to mask $F$ contains (\ie the statements having at least 11 tokens). This may end up in generating duplicates due to different functions from which we masked the only part being different, thus resulting in duplicates. For this reason, we perform a second deduplication round on all the datasets. %,  ending up with \textcolor{red}{xxx} instances for the \emph{short} set, \textcolor{red}{xxx} instances for the \emph{medium} set and \textcolor{red}{xxx} instances for the \emph{long} set.

The final set of (masked) functions is then flattened to obtain the model input by replacing all new line characters (\ie `\texttt{\textbackslash n}’) with a special tag, ⟨NEW\_LINE⟩, and remove all tabs (\ie `\texttt{\textbackslash t}’) and white spaces used to indent the code. We randomly split each set (\emph{short}, \emph{medium}, \emph{long}) into training (80\%), validation (10\%) and test (10\%) by making sure that all the instances obtained from the same function fall into the same set.

Finally, for each of the three datasets (\emph{short}, \emph{medium}, \emph{long}), we limit the number of training instances to 280k and, proportionally, those of the test and evaluation set to 35k. This is done to reduce the training cost of our study, as explained later, required to train and test 32 DL models. These numbers (\ie 280k and 35k) are inherited from the (smaller) Python dataset that we will describe in the next section. In other words, we aligned the size of the Java and of the Python datasets towards the smallest one (Python), due to our computational resources.

To summarize, at the end of this process we have three datasets (\emph{short}, \emph{medium}, and \emph{long}) each split into training, validation, and test, all containing the same number of instances having, however, different lengths.

We also built a fourth dataset, named \emph{mixed}, consisting of a mix of the three lengths: 1/3 of instances comes from the \emph{short} dataset, 1/3 from the \emph{medium}, and 1/3 from the \emph{long}. In this case we only built the training and the validation sets, since we will test the model trained on the mixed dataset on the \emph{short}, \emph{medium} and \emph{long} test sets.

\subsubsection{Python dataset: block-level code completion task}
Similarly to what we discussed for Java, we build three Python datasets (\emph{short}, \emph{medium}, and \emph{long}) featuring instances (functions) having different lengths but characterized by the same task complexity (\ie same number of masked tokens to predict). The main difference between the Java and the Python dataset is that the latter simulates block-level completion, thus possibly featuring completions spanning across multiple statements. 
The process used to build the Python datasets resembles the one we presented for Java. Thus, we only briefly summarize it here. We removed all functions not containing any code block (2,833,017). For consistency with the Java datasets, we decided to keep the same task complexity, meaning that we target the masking of the last 11 Python tokens within a given block. Thus, we remove from the dataset all functions not containing any valid block to mask (\ie a block consisting of at least 11 tokens).

We then sort the remaining functions by their length and split them into three sets of the same size: \emph{short} (featuring functions with length from 30 to 150), \emph{medium} (from 151 to 309), and \emph{long} (from 310 to 1024). For each function $F$ we created $n$ instances, each one having a different block featuring its last 11 tokens masked. For the same reasons previously explained, we remove any duplicates created at this stage.
 
We replace all characters used to indent the code (\ie `\texttt{\textbackslash n}’, `\texttt{\textbackslash t}’ and extra white spaces) with a special tag: \texttt{⟨TAB⟩}. This allows to flatten each function without losing information about the indentation, which is fundamental for the Python syntax. 

Finally, we split each dataset (\emph{short}, \emph{medium}, \emph{long}) into training, validation, and test using the same procedure described for Java. For each dataset, the training contains 280k instances, while the evaluation and test contain 35k instances. These numbers have been dictated by the smallest dataset involved in our study, being the \emph{short} Python dataset. Aligning the size of all datasets removes another possible confounding factor.

\subsection{Data Collection \& Analysis}\label{subsec:experiments}

\input{tabs/hyperparams}

To answer our research question, we train eight models (four per each of the subject languages, namely Python and Java) for each of the four experimented position encoding schemes (\ie Sinusoidal, xPOS, ALiBi, and T5). This leads to a total of 32 trained models. The four models for each language have been trained on datasets featuring code completions having inputs (\ie the Java or Python function to complete) characterized by different lengths (\ie \emph{short}, \emph{medium}, \emph{long}, and \emph{mix}). Then, each of these models have been used to generate predictions on three test sets featuring code completions of different lengths (\ie \emph{short}, \emph{medium}, and \emph{long}). 

This allows us to verify if, for example, a model trained on \emph{short} code completions can generalize to a test set containing \emph{long} instances. Also, we can verify whether a model trained on code completions having a mixture of lengths (\ie featuring short, medium, and long sequences) can achieve on each of the three test sets (\ie \emph{short}, \emph{medium}, and \emph{long}) results competitive with those of models specialized (\ie trained only) on instances having a specific length. For example, we can check whether the model trained on the mixture of lengths achieves on the \emph{short} test set performance comparable to those of the model trained on the \emph{short} training set. Remember that the amount of instances in each training set is fixed. Thus, observed differences should be due to the length of the employed training instances.

To reduce confounding factors, we used the same hyperparameters amongst all 32 models. The adopted hyperparameters are those suggested in the paper originally proposing the Transformer architecture \cite{vaswani2017attention} and are reported in Table \ref{tab:hyperparams}. This design decision also avoided the need for an expensive hyperparameters tuning involving 32 different models.

All models have been trained with the Adam optimizer \cite{kingma2014adam} with a cosine learning rate scheduler using a warmup of 2,000 steps. We used a vocabulary size of 50k for the tokenizer, which was shared across all the models.

For implementing and training the Transformers, we used x-transformers \cite{xtransformers} and Pytorch Lightning \cite{Falcon_PyTorch_Lightning_2019}. Additionally, when generating samples, we used Nucleus Sampling \cite{holtzman2019curious} with a $top_p = 0.95$ and stopped generations once the \texttt{⟨EOS⟩} token was produced or the maximum number of tokens, $128$, were produced. We trained all models for a maximum of five epochs and used the best performing checkpoint based on validation loss, which happened to always be the models trained on all five epochs. While this suggests that better performance might be obtained with additional training epochs, it is important to note that our goal is not to train the most performant code completion model, but rather to show that, given a certain training cost, the consistency of training and test data (in terms of predictions length) substantially impacts performance at inference time. 

% \textcolor{red}{Nathan, here we need to say for how much we trained the model. I hope we used an early stopping or something exploiting the evalaution set otherwise there's no reason for the evaluation set to exist since we don't do hyperparameters tuning. Can you add this part here?}\nathan{I had us saving the best checkpoint, but looking back at it, it looks like I do not use it for training. I only use the latest model. All mixed models had the best checkpoint used.}

To assess the performance of the models on the test sets, we collect the predictions they generate and measure the percentage of Exact Match (EM) with respect to the expected target. An EM indicates that the code generated by the model for the completion instance is identical to the target. 

We also compute metrics usually adopted in the assessment of generative models, namely the BLEU \cite{papineni2002bleu}, ChrF \cite{popovic2015chrf}, Levenshtein Distance \cite{levenshtein1966binary}, ROUGE \cite{lin2004rouge}, and METEOR \cite{banerjee2005meteor} score with respect to the target.

BLEU  is a popular automatic metric for machine translation tasks due to the high correlation to human judgement. It has become a standard metric in code completion tasks \cite{lu2021codexglue} since it measures the overlap of a predicted sequence and a set of reference sequences in terms of n-grams. ChrF is a character level metric which averages the F-score of 1 to 6-grams of characters. Levenshtein Distance is a measure of the minimal edit operations (\ie insert, modify, and remove), that would be needed to convert the predicted sequence into the target one, and it has been used in assessing the models' performance in previous code completion studies \cite{ciniselli2021empirical}. ROUGE, and specifically RougeL, is a metric that measures the longest common subsequence between the predicted and ground truth sequences. Lastly, Meteor is also an F-score, where the recall is weighted nine times more than the precision. 

Additionally, predictions are penalized for not having adjacent unigrams that exist in the ground truth. We also measure the Cross-Entropy of the generated predictions (\ie a measure of the surprise of the model when predicting the ground truth sequence).

While we computed all the above-described metrics, we only discuss the results achieved in terms of EM, ChrF, and RougeL. The former (EM) is an easy-to-interpret proxy of the model's performance. ChrF and RougeL, instead, have been found to be best at measuring performance compared to human evaluation and allow to claim significance (95\% confidence) if the difference between two models on code generation tasks is greater than two points \cite{evtikhiev2022out}. Our full analysis can be found in our replication package \cite{replication}. For implementing these metrics, we used the Huggingface's datasets library \cite{lhoest-etal-2021-datasets}, which contains a large selection of automated metrics for the evaluation of generative models.

%% file: tabs/hyperparams.tex
\begin{table}[]
\centering
\caption{Hyperparameters used and searched.}
\label{tab:hyperparams}
\begin{tabular}{l|r}
\hline
\textbf{Hyperparameter} & \textbf{Values} \\ \hline
Learning Rate           & 1e-4            \\
Batch Size              & 256             \\
Inner Dimension         & 512             \\
Encoder Max Length      & 1,024            \\
Encoder Layers          & 6               \\
Encoder Heads           & 8               \\
Decoder Max Length      & 128             \\
Decoder Layers          & 8               \\
Decoder Heads           & 6               \\ \hline
\end{tabular}
\end{table}

%% file: texs/05_results.tex
% !TEX root = ../main.tex
\section{Results Discussion}\label{sec:results}

\input{tabs/exact_results}
\input{tabs/chrf_results}
\input{tabs/rouge_results}

Tables \ref{tab:exact_results}, \ref{tab:chrf_results}, and \ref{tab:rougel_results} show the results in terms of EM, ChrF, and RougeL, respectively, achieved by the four positional encoding schemes when trained on datasets featuring code completions of different lengths (columns) and tested on the \emph{short}, \emph{medium}, and \emph{long} test sets (rows). 
The results are reported for both Java and Python. To provide a concrete example, let us consider the EM results reported in the Table \ref{tab:exact_results}. Here, the Sinusoidal schema trained on \emph{short} Java completions generated 10.81\% EM predictions when tested on \emph{short} instances (\ie those resembling the training instances). 

Instead, when the training is performed on instances having a \emph{medium} length, the percentage of EM predictions drops, on the same \emph{short} test set, to 2.91\%, finally falling at 0.50\% when the training was performed on \emph{long} instances. 

Similar results are observed for Python in which, however, the percentage of EM predictions is lower, moving from 1.57\% achieved on the \emph{short} test set when training on \emph{short} instances down to the 0.03\% when tested on the \emph{long} ones.

Tables \ref{tab:exact_results}, \ref{tab:chrf_results}, and \ref{tab:rougel_results} also contain two ``Avg. $\Delta$'' columns (one per language). Given a row in one of the tables (\eg the first row in Table \ref{tab:exact_results} reporting the performance of the Sinusoidal schema when run on the \emph{short} test set), the Avg. $\Delta$ indicates the relative change in performance observed, on average, for the models trained on different lengths (in our example, those trained on \emph{medium} and \emph{long} completions) when compared the one specialized on lengths related to that row (\ie \emph{short}). Indeed, the 84.22\% shown as average $\Delta$ in the subject row is the result of:

$$
\frac{\frac{10.81\% - 2.91\%}{10.81\%} + \frac{10.81\% - 0.50\%}{10.81}}{2} = 84.22\%
$$

where 10.81\% is the percentage of EM predictions for Java generated by the Sinusoidal schema when trained and tested on code completions of \emph{short} length, while 2.91\% and 0.50\% are the EM scores achieved by the Sinusoidal schema when trained on \emph{medium} and \emph{long} instances, respectively, while still being tested on \emph{short} instances.

Finally, Tables \ref{tab:exact_results}, \ref{tab:chrf_results}, and \ref{tab:rougel_results} adopt three styles to highlight findings in the context of a specific test set length. Let us focus on the Java results achieved on the \emph{short} test set in terms of EM (Table \ref{tab:exact_results}). The black box shows the best-performing combination of $\langle$\emph{encoding schema}, \emph{training length}$\rangle$ for such a test set (\ie T5 trained on \emph{short} completions). The bold values highlight, for each encoding schema, the best-performing training length for such a test set (\ie in all cases, training on \emph{short} instances works better when testing on \emph{short} instances). The red value in each ``Avg. $\Delta$'' column highlights, instead, the encoding schema manifesting the lowest relative drop in performances when moving from a training length matching the instances in the test set (\eg training on \emph{short}, testing on \emph{short}) to the other training lengths. In this case, the lowest relative drop in terms of EM predictions is exhibited by xPOS. Note that a lowest relative drop indicates a better ability of the encoding schema to generalize to unseen lengths. 

The first observation that can be made from the three tables is that T5's positional encoding schema performs better than all other approaches. Such a finding is consistently captured by all metrics, including ChfR and RougeL for which the difference is always substantially higher than two points, indicating a statistically significant difference at 95\% confidence \cite{evtikhiev2022out}. Being the best performing one, however, does not save T5 from a strong general observation that can be made across the board for all training schemes: They all suffer from a major degradation in performance when applied on code completions having a length different from the one they have been trained on. Interestingly, the degradation is not only observed when the models are tested on instances being longer (likely more complex to handle) than those they have been trained on, but also in the opposite direction. This can be easily seen in Tables \ref{tab:exact_results}, \ref{tab:chrf_results}, and \ref{tab:rougel_results} by the fact that (i) the average $\Delta$ values are always positive, and (ii) the bold values in a given test set length are always associated with the same length in the training set.

The second observation concerns the encoding schema reporting the lowest average drop in performance (red values in the ``Avg. $\Delta$'' column in the three tables). Overall, also from this perspective, T5 seems to be the best choice. There are a few exceptions to this trend, depending on the test set under analysis and on the metric used as proxy for performance. 

For example, on the \emph{short} Java test set, T5 is the second best in class in terms of EM and ChfR score, while confirming its leadership when looking at the RougeL score. By considering all 18 combinations of test set length (3), language (2), and evaluation metrics (3), T5 is the one exhibiting the lowest relative drop in 11 (61\%) of cases, and the second-best in additional 3 cases (17\%). Still, as observed, T5 also exhibits major drops in performance when working on sequence lengths unseen during training. For example, there is an absolute drop of 13.41\% in terms of EM predictions when testing T5 on the \emph{long} dataset when trained on \emph{short} sequences as compared to the one trained on \emph{long} sequences (17.03\% \emph{vs} 3.62\%). The trend is confirmed when looking at the ChfR and the RougeL scores.

xPOS is the second best performing schema, both in terms of absolute performance and generalization to different lengths. ALiBi and Sonusoidal follow, exhibiting similar performances from both perspectives.

\begin{resultbox}
\textbf{Take Away \#1:} T5's positional encoding scheme achieves the best overall performance across metrics, lengths, and languages. Also, it is also better at generalizing to unseen lengths. In general, however, all encoding schemes suffer generalization issues for unseen lengths.  
\end{resultbox}

\textbf{Differences across languages.} Overall, our main findings hold on both languages. These include: (i) the lack of generalizability to unseen lengths of any of the experimented encoding schemes; and (ii) the superiority of T5 both in terms of absolute performance and relative drop when dealing with unseen lengths. 

We do not compare the performance achieved on the two languages since (i) the test sets are different, (ii) the code completion tasks are different (statement-level \emph{vs.} block-level), and (iii) the syntaxt of the two languages makes the prediction tasks quite different, since Python requires the generation of the \texttt{<TAB>} indentation tokens while Java does not.

\begin{resultbox}
\textbf{Take Away \#2:} On both languages, all encoding schemes struggle to generalize to unseen lengths. T5 confirms its superiority on both Java and Python code.
\end{resultbox}

\input{tabs/exact_mix}

\input{tabs/chrf_mix}
\input{tabs/rouge_mix}

\textbf{Impact of training diversity.} Tables \ref{tab:exact_mix} (EM), \ref{tab:chrf_mix} (ChrF), and \ref{tab:rouge_mix} (RougeL) report the results achieved by the four encoding schemas (rows) when trained on the \emph{mix} dataset (\ie the one featuring a mixture of instances taken from the \emph{short}, \emph{medium}, and \emph{long} datasets)  and tested on the three datasets featuring sequences of different length (columns). Note that the \emph{mix} training dataset has exactly the same number of instances of the other length-specific datasets, thus do not introducing a confounding variable related to the training size. 

The $\Delta$ associated to each combination of encoding schema and test set is the relative change in performance with respect to the same schema exclusively trained for sequences of the corresponding length. For example, in terms of ChrF score (Table \ref{tab:chrf_mix}), T5 trained on \emph{short} Java sequences achieves a 42.97\% ChrF score when tested on the \emph{short} dataset. Such a score grows to 45.47\% when T5 is trained on the \emph{mix} dataset, with a relative improvement equal to (45.47\% - 42.97\%)/42.97\% = +5.82\%. The achieved findings confirm the superiority of T5 in this scenario as well. 

Most importantly, we found that relying on a mixture of lengths during training is generally sufficient to achieve results approaching, and in some cases improving, than those achieved by specifically training the model for the target sequence length. Indeed, by comparing the relative $\Delta$ reported, for example, in Table \ref{tab:chrf_mix} (ChfR scores when training on the \emph{mix} dataset) to those reported in Table \ref{tab:chrf_results} (ChfR scores when training on datasets featuring functions having different lengths), it is possible to observe a major difference in terms of magnitude of the deltas, with those in \ref{tab:chrf_mix} being substantially smaller. 

\eject

This indicates that, while in some cases training on a specific length range $l$ could help in achieving better performances on test instances fitting $l$, training on a mixture of lengths is a safe choice, since it would not result in dramatic lost of performances as those observed in Table \ref{tab:chrf_results}. 

Finally, it is worth mentioning that while training diversity was very successful in Java, it helped less in Python. Still, when comparing the deltas in performance between Tables \ref{tab:exact_results}, \ref{tab:chrf_results}, \ref{tab:rougel_results} (\ie the delta in performance when testing on prediction lengths different from the one the model has been trained for \emph{vs} testing on the same prediction length) with the deltas in Tables \ref{tab:exact_mix}, \ref{tab:chrf_results}, and \ref{tab:rouge_mix} (\ie the delta in performance when exploiting training diversity \emph{vs} when training on the same prediction length used for testing) it can be seen that even for Python training diversity substantially helped.

\begin{resultbox}
\textbf{Take Away \#3:} Training on a mixture of lengths being representative of those that will be seen during testing but also including other types of lengths might be the safest choice in most of cases. Only in scenarios in which even minor increases in performance are considered valuable, experimenting with a combination of models specialized on different lengths might be worthwhile, to then decide the best strategy to adopt.
\end{resultbox}

%% file: tabs/exact_results.tex
\begin{table*}[]
\centering
\caption{Exact Match Score ($\uparrow$) achieved by the different position encoding schemes.\vspace{-0.3cm}}
\label{tab:exact_results}
\begin{tabular}{cc|cccc|cccc}
\hline
\multirow{2}{*}{\textbf{Test Set}} & \multirow{2}{*}{\textbf{Encoding Schema}}        & \multicolumn{4}{c|}{\textbf{Java}}                                  & \multicolumn{4}{c}{\textbf{Python}}    \\
\multicolumn{2}{c|}{}                                  & \textbf{Short} & \textbf{Medium} & \textbf{Long} & \textbf{Avg. $\Delta$}                      & \textbf{Short} & \textbf{Medium} & \textbf{Long} & \textbf{Avg. $\Delta$}   \\ \hline
\multirow{4}{*}{\textbf{Short}}  & \textbf{Sinusoidal} & \textbf{10.81\%}        & 2.91\%          & 0.50\%        & 84.23\%                    & \textbf{1.57\%}         & 0.26\%          & 0.03\%        & 90.76\%  \\
                                 & \textbf{xPOS}     & \textbf{12.49\%}        & 4.87\%          & 0.85\%        & \textcolor{red}{77.10\%}   & \textbf{2.33\%}         & 0.64\%          & 0.07\%        & 84.76\%  \\
                                 & \textbf{ALiBi}      & \textbf{10.97\%}        & 3.09\%          & 0.36\%        & 84.28\%                    & \textbf{1.48\%}         & 0.26\%          & 0.03\%        & 90.20\%  \\
                                 & \textbf{T5}         & \cellcolor{black}\textbf{\textcolor{white}{18.39\%}} 
                                                                                 & 6.66\%          & 1.38\%        & 78.14\%                    & \cellcolor{black}\textbf{\textcolor{white}{5.77\%}} 
                                                                                                                                                                          & 3.19\%          & 1.51\%        & \textcolor{red}{59.27\%} \\ \hline
                                 
\multirow{4}{*}{\textbf{Medium}} & \textbf{Sinusoidal} & 1.61\%   & \textbf{6.67\%}   & 3.32\%        & 63.04\%            & 0.57\%         & \textbf{1.33\%}          & 0.72\%        & 51.50\%  \\
                                 & \textbf{xPOS}     & 2.85\%   & \textbf{9.36\%}   & 8.02\%        & 41.93\%            & 1.53\%         & \textbf{2.84\%}          & 1.73\%        & 42.61\%  \\
                                 & \textbf{ALiBi}      & 1.56\%   & \textbf{6.61\%}   & 3.68\%        & 60.36\%            & 0.49\%         & \textbf{1.36\%}          & 0.64\%        & 58.46\%  \\
                                 & \textbf{T5}         & 5.29\%   & \cellcolor{black}\textbf{\textcolor{white}{14.06\%}}
                                                                                     & 11.33\%  & \textcolor{red}{40.90\%} & 3.16\%         & \cellcolor{black}\textbf{\textcolor{white}{5.71\%}}
                                                                                                                                                                            & 5.99\%        & \textcolor{red}{19.88\%} \\ \hline
                                 
\multirow{4}{*}{\textbf{Long}}   & \textbf{Sinusoidal} & 0.60\%  & 2.85\%  & \textbf{8.27\%}   & 79.14\%           & 0.05\%         & 0.61\%   & \textbf{8.81\%}        & 96.25\%          \\
                                 & \textbf{xPOS}     & 1.57\%  & 5.47\%  & \textbf{11.33\%}  & 68.93\%           & 0.48\%         & 1.80\%   & \textbf{11.34\%}       & 89.95\%          \\
                                 & \textbf{ALiBi}      & 0.57\%  & 2.79\%  & \textbf{8.29\%}   & 79.73\%           & 0.08\%         & 0.50\%   & \textbf{8.84\%}        & 96.72\%          \\
                                 & \textbf{T5}         & 3.62\%  & 9.59\%  & \cellcolor{black}\textbf{\textcolor{white}{17.03\%}}       
                                                                                     & \textcolor{red}{61.22\%}    & 3.10\%   & 7.19\%   & \cellcolor{black}\textbf{\textcolor{white}{20.73\%}} & \textcolor{red}{75.18\%}
\end{tabular}
\vspace{0.3cm}
\end{table*}

%% file: tabs/chrf_results.tex
\begin{table*}[]
\centering
\caption{ChrF Score ($\uparrow$) achieved by the different position encoding schemes.\vspace{-0.3cm}}
\label{tab:chrf_results}
\begin{tabular}{cc|cccc|cccc}
\hline
\multirow{2}{*}{\textbf{Test Set}} & \multirow{2}{*}{\textbf{Encoding Schema}}  & \multicolumn{4}{c|}{\textbf{Java}}                          & \multicolumn{4}{c}{\textbf{Python}}                     \\
\multicolumn{2}{c|}{}                                    & \textbf{Short}   & \textbf{Medium}  & \textbf{Long}    & \textbf{Avg. $\Delta$}            & \textbf{Short}   & \textbf{Medium}  & \textbf{Long}    & \textbf{Avg. $\Delta$}\\ \hline
\multirow{4}{*}{\textbf{Short}}   & \textbf{Sinusoidal}  & \textbf{30.72\%} & 17.70\% & 13.19\% & 49.72\%                                     & \textbf{37.48\%} & 32.56\% & 30.00\% & 16.54\%  \\
                         & \textbf{xPOS}               & \textbf{34.57\%} & 23.36\% & 16.89\% & \textcolor{red}{41.78\%}                                   & \textbf{41.83\%} & 37.51\% & 33.06\% & 15.65\%  \\
                         & \textbf{ALiBi}                & \textbf{30.95\%} & 17.46\% & 13.02\% & 50.76\%                                     & \textbf{37.50\%} & 33.10\% & 30.31\% & 15.45\%  \\
                         & \textbf{T5} & \cellcolor{black}\textbf{\textcolor{white}{42.97\%}} & 27.43\%  & 17.37\% & 47.87\% & \cellcolor{black}\textbf{\textcolor{white}{50.86\%}} & 48.41\% & 43.30\% & \textcolor{red}{9.84\%}\\ \hline

\multirow{4}{*}{\textbf{Medium}}  & \textbf{Sinusoidal}  & 16.57\% & \textbf{25.01\%} & 19.94\% & 27.01\%                                     & 35.99\% & \textbf{37.17\%} & 35.21\% & \textcolor{red}{4.22\%}  \\
                         & \textbf{xPOS}               & 21.41\% & \textbf{30.96\%} & 27.68\% & 20.72\%                                     & 40.84\% & \textbf{43.95\%} & 39.30\% & 8.83\%  \\
                         & \textbf{ALiBi}                & 16.39\% & \textbf{24.93\%} & 20.29\% & 26.43\%                                     & 35.65\% & \textbf{36.88\%} & 34.67\% & 4.66\%   \\
                         & \textbf{T5} & 29.18\% & \cellcolor{black}\textbf{\textcolor{white}{39.67\%}} & 35.95\%  & \textcolor{red}{17.91\%} & 47.45\% & \cellcolor{black}\textbf{\textcolor{white}{52.61\%}} & 51.36\% & 6.09\% \\ \hline

\multirow{4}{*}{\textbf{Long}}    & \textbf{Sinusoidal}  & 13.93\% & 17.19\% & \textbf{25.03\%} & 37.83\%                                     & 32.79\% & 34.72\% & \textbf{41.71\%} & \textcolor{red}{19.07\%}          \\
                         & \textbf{xPOS}               & 18.30\% & 25.18\% & \textbf{32.38\%} & 32.86\%                                     & 38.36\% & 42.00\% & \textbf{50.58\%} & 20.56\%          \\
                         & \textbf{ALiBi}                & 13.47\% & 17.24\% & \textbf{24.91\%} & 38.36\%                                     & 32.36\% & 34.50\% & \textbf{41.66\%} & 19.76\%           \\
                         & \textbf{T5} & 27.14\% & 36.42\% & \cellcolor{black}\textbf{\textcolor{white}{44.93\%}} & \textcolor{red}{29.27\%}  & 48.14\% & 54.86\% & \cellcolor{black}\textbf{\textcolor{white}{64.96\%}} & 20.72\%
\end{tabular}
\vspace{0.3cm}
\end{table*}

%% file: tabs/rouge_results.tex
\begin{table*}[]
\centering
\caption{ROUGE-L Score ($\uparrow$) achieved by the different position encoding schemes.\vspace{-0.3cm}}
\label{tab:rougel_results}
\begin{tabular}{cc|cccc|cccc}
\hline

\multirow{3}{*}{\textbf{Test Set}} & \multirow{3}{*}{\textbf{Encoding Schema}}      & \multicolumn{4}{c|}{\textbf{Java}}                 & \multicolumn{4}{c}{\textbf{Python}}\\
\multicolumn{2}{c|}{}                      & \textbf{Short} & \textbf{Medium} & \textbf{Long} & \textbf{Avg. $\Delta$}                    & \textbf{Short}     & \textbf{Medium} & \textbf{Long} & \textbf{Avg. $\Delta$}\\ \hline
\multirow{4}{*}{\textbf{Short}}  & \textbf{Sinusoidal}  & \textbf{38.80\%}   & 24.56\%     & 17.68\%   & 45.57\%                  & \textbf{39.39\%}   & 33.61\%         & 30.16\%       & 19.05\% \\
                                 & \textbf{xPOS}     & \textbf{45.61\%}   & 35.23\%     & 25.66\%   & \textcolor{red}{33.25\%}               & \textbf{45.01\%}   & 40.10\%         & 32.96\%       & 18.84\% \\
                                 & \textbf{ALiBi}      & \textbf{39.24\%}   & 24.71\%     & 17.85\%   & 45.77\%                                              & \textbf{39.41\%}   & 33.81\%         & 30.62\%       & 18.26\% \\
                                 & \textbf{T5}         & \cellcolor{black}\textbf{\textcolor{white}{52.47\%}}  
                                                                            & 38.99\%     & 25.92\%   & 38.15\%                   & \cellcolor{black}\textbf{\textcolor{white}{52.84\%}}   & 50.46\%   & 44.92\%  & \textcolor{red}{9.75\%}\\ \hline

\multirow{4}{*}{\textbf{Medium}} & \textbf{Sinusoidal} & 22.08\%    & \textbf{29.63\%}    & 23.63\%    & 22.87\%                   & 36.26\%        & \textbf{37.15\%}         & 33.82\%       & 5.68\% \\
                                 & \textbf{xPOS}     & 30.98\%    & \textbf{41.28\%}    & 36.71\%    & 18.01\%                   & 43.14\%        & \textbf{46.43\%}         & 41.28\%       & 9.09\% \\
                                 & \textbf{ALiBi}      & 21.39\%    & \textbf{29.57\%}    & 24.18\%    & 22.95\%                   & 35.98\%        & \textbf{36.74\%}         & 33.81\%       & \textcolor{red}{5.02\%}\\
                                 & \textbf{T5}         & 39.12\%    & \cellcolor{black}\textbf{\textcolor{white}{49.36\%}}    
                                                                                          & 45.66\%    & \textcolor{red}{14.12\%}                            & 49.04\%        & \cellcolor{black}\textbf{\textcolor{white}{54.49\%}} & 53.14\% & 6.24\% \\ \hline

\multirow{4}{*}{\textbf{Long}}   & \textbf{Sinusoidal} & 19.05\%    & 22.27\%     & \textbf{26.60\%}   & 22.33\%                   & 32.80\%        & 34.07\%         & \textbf{39.83\%}       & \textcolor{red}{16.06\%}\\
                                 & \textbf{xPOS}     & 26.79\%    & 36.05\%     & \textbf{41.01\%}   & 23.38\%                   & 41.17\%        & 44.39\%         & \textbf{51.65\%}       & 17.17\% \\
                                 & \textbf{ALiBi}      & 18.69\%    & 22.33\%     & \textbf{26.66\%}   & 23.07\%                   & 32.70\%        & 33.90\%         & \textbf{39.80\%}       & 16.33\% \\
                                 & \textbf{T5}         & 36.20\%    & 46.35\%     & \cellcolor{black}\textbf{\textcolor{white}{52.84\%}}   
                                                                                                       & \textcolor{red}{21.89\%}  & 49.71\%        & 56.35\%         & \cellcolor{black}\textbf{\textcolor{white}{65.52\%}}  & 19.06\%
\end{tabular}
\vspace{0.2cm}
\end{table*}

%% file: tabs/exact_mix.tex
\begin{table*}[]
\centering
\caption{Exact Match Mix ($\uparrow$) achieved by the different position encoding schemes.\vspace{-0.3cm}}
\label{tab:exact_mix}
\begin{tabular}{c|p{1.5cm}p{1.5cm}p{1.5cm}|p{1.5cm}p{1.5cm}p{1.5cm}}
\hline
\multirow{2}{*}{}   & \multicolumn{3}{c|}{\textbf{Java - Test Set}}               & \multicolumn{3}{c}{\textbf{Python - Test Set}}              \\
                    & \textbf{Short} & \textbf{Medium} & \textbf{Long} & \textbf{Short} & \textbf{Medium} & \textbf{Long} \\ \hline
\textbf{Sinusoidal} & 9.85\%         & 7.05\%          & 8.81\%        & 1.03\%         & 1.04\%          & 7.50\%        \\
\textbf{$\Delta$}      & -8.88\%         & +5.70\%         & +6.53\%       & -34.39\%         &-21.80\%          & -14.87\%        \\ \hline
\textbf{xPOS}     & 11.99\%        & 9.73\%          & 11.42\%       & 2.49\%         & 2.76\%          & 10.67\%       \\
\textbf{$\Delta$}      & -4.00\%         & +3.95\%         & +0.79\%       & +6.87\%        & -2.82\%          & -5.91\%        \\ \hline
\textbf{ALiBi}      & 10.12\%        & 6.85\%          & 8.53\%        & 1.00\%         & 0.98\%          & 7.34\%        \\
\textbf{$\Delta$}      & -7.75\%         & +3.63\%         & +2.90\%       & -32.43\%         & -27.94\%          & -16.97\%        \\ \hline
\textbf{T5}         & 19.11\%        & 17.57\%         & 18.63\%       & 3.19\%         & 3.69\%          & 12.22\%       \\
\textbf{$\Delta$}      & +3.92\%        & +24.96\%         & +9.40\%       & -44.71\%         & -35.38\%          & -41.05\%       
\end{tabular}
%\vspace{0.4cm}
\end{table*}

%% file: tabs/chrf_mix.tex
\begin{table*}[]
\centering
\caption{ChrF Mix ($\uparrow$) achieved by the different position encoding schemes.\vspace{-0.3cm}}
\label{tab:chrf_mix}
\begin{tabular}{c|p{1.5cm}p{1.5cm}p{1.5cm}|p{1.5cm}p{1.5cm}p{1.5cm}}
\hline
\multirow{2}{*}{}   & \multicolumn{3}{c|}{\textbf{Java - Test Set}}               & \multicolumn{3}{c}{\textbf{Python - Test Set}}              \\
                    & \textbf{Short} & \textbf{Medium} & \textbf{Long} & \textbf{Short} & \textbf{Medium} & \textbf{Long} \\ \hline
\textbf{Sinusoidal} & 30.49\%        & 27.04\%           & 26.60\%         & 36.48\%    & 36.24\%         & 40.91\%   \\
\textbf{$\Delta$}   & -0.75\%        & +8.12\%           & +6.27\%         & -2.67\%    & -2.50\%         & -1.92\%   \\ \hline
\textbf{xPOS}     & 34.97\%        & 33.20\%           & 33.34\%         & 43.83\%    & 45.07\%         & 51.33\%   \\
\textbf{$\Delta$}   & +1.16\%        & +7.24\%           & +2.96\%         & +4.78\%    & +2.55\%         & +1.48\%   \\ \hline
\textbf{ALiBi}      & 30.42\%        & 26.72\%           & 26.28\%         & 36.44\%    & 36.14\%         & 40.93\%   \\
\textbf{$\Delta$}   & -1.71\%        & +7.18\%           & +5.50\%         & -2.83\%    & -2.01\%         & -1.75\%   \\ \hline
\textbf{T5}         & 45.47\%        & 46.11\%           & 47.73\%         & 45.94\%    & 48.33\%         & 55.00\%   \\
\textbf{$\Delta$}   & +5.82\%        & +16.23\%          & +6.23\%         & -9.67\%    & -8.14\%         & -15.33\%         
\end{tabular}
%\vspace{0.4cm}
\end{table*}

%% file: tabs/rouge_mix.tex
\begin{table*}[]
\centering
\caption{ROUGE-L Mix ($\uparrow$) achieved by the different position encoding schemes.\vspace{-0.3cm}}
\label{tab:rouge_mix}
\begin{tabular}{c|p{1.5cm}p{1.5cm}p{1.5cm}|p{1.5cm}p{1.5cm}p{1.5cm}}
\hline
\multirow{2}{*}{}   & \multicolumn{3}{c|}{\textbf{Java - Test Set}}               & \multicolumn{3}{c}{\textbf{Python - Test Set}}              \\
                    & \textbf{Short} & \textbf{Medium} & \textbf{Long} & \textbf{Short} & \textbf{Medium} & \textbf{Long} \\ \hline
\textbf{Sinusoidal} & 36.85\%        & 31.54\%         & 29.37\%       & 37.85\%        & 36.18\%         & 39.35\%    \\
\textbf{$\Delta$}      & -5.03\%        & +6.45\%         & +10.41\%      & -3.91\%        & -2.61\%         & -1.21\%    \\ \hline
\textbf{xPOS}     & 45.26\%        & 43.36\%         & 42.60\%       & 46.27\%        & 47.33\%         & 52.25\%    \\
\textbf{$\Delta$}      & -0.77\%        & +5.04\%         & +3.88\%       & +1.94\%        & +1.94\%         & +1.16\%    \\ \hline
\textbf{ALiBi}      & 36.83\%        & 31.03\%         & 28.82\%       & 37.80\%        & 36.10\%         & 39.28\%    \\
\textbf{$\Delta$}      & -6.14\%        & +4.94\%         & +8.10\%       & -1.74\%        & -1.74\%         & -1.31\%    \\ \hline
\textbf{T5}         & 52.60\%        & 53.96\%         & 55.18\%       & 48.20\%        & 50.47\%         & 55.94\%    \\
\textbf{$\Delta$}      & +0.25\%        & +9.32\%         & +4.43\%       & -8.78\%        & -7.38\%         & -14.62\%      
\end{tabular}
%\vspace{0.4cm}
\end{table*}

%% file: texs/06_threats.tex
% !TEX root = ../main.tex

\section{Threats to Validity \& Future Work}\label{sec:threats}

\noindent\textbf{Threats to Internal Validity.} In order to control for various levels of bias that can creep into our evaluation, we ensured to hold as many variables as possible in our datasets and models constant. This involved ensuring that there were no duplicates across the different training splits, both in the input and target \cite{allamanis2019adverse}. Also, we held constant the hyperparameters across our different models and only changed the type of length they were trained on. However, despite these thorough mitigation strategies, bias can still be present in our empirical study. 

% One potential for bias being introduced into our study is our decision to perform a hyperparameter search on the T5 model and simple complexity type. This may have biased our results to favour the T5 model due to selection of optimal hyperparameters that might not work well for ALiBi.\medskip

\noindent\textbf{Threats to Construct Validity.} To mitigate threats to construct validity, we calculated a range of different metrics that have been commonly used in code completion literature. Additionally, we focus our discussion either on metrics that  have been shown to correlate with human preference and that are more statistically stable \cite{evtikhiev2022out} (\ie ChfR and RougeL) or that allow for a simple interpretation such as EM.

While there have been new recent metrics that are specific to code data for code completion, namely CodeBLEU \cite{ren2020codebleu} and functional-correctness \cite{hendrycks2021measuring, chen2021evaluating}, we did not compute these for the following reasons. CodeBLEU has been shown to be not as stable as ChrF and RougeL  \cite{evtikhiev2022out}. Unfortunately, functional-correctness was not even an option for our evaluation due to the lack of unit tests for our test examples. 

Besides the metric used, the type of code completion performed can result in bias as there has been some studies showing that synthetic benchmarks of code completion where the completions are randomized do not necessarily reflect the performance of real-world code completions \cite{hellendoorn2019code}. 

% Lastly, our choice to not have our models predict an End of Sequence (EOS) token does influence our measurements. Specifically, our measurements should be thought of as an optimistic performance given an oracle on the length of what the code completion should be, as discussed by Newman \etal \cite{newman2020eos}. Such a design choice does not invalidate our results as we are concerned with generalizability, not measuring end of sequence prediction performance.\medskip %Additionally, we do not believe training without an EOS token is an unrealistic setting since other, ad-hoc, post-processing of the generated code completion can be done to align it with the incomplete code.

% Talk about EOS impacting metric performance

\noindent\textbf{Threats to External Validity.} We investigated two Transformer architectures to mitigate the threats to external validity. Additionally, we measured multiple types of metrics and constructed our datasets in such a way as to hopefully mimic realistic code completion scenarios. Additionally, we used two popular programming languages, namely Java and Python, to better ensure our results generalize across languages.

% However, our datasets focused only on the Python programming language and the task of code completion. Our results of the generalizability to unseen complexities and lengths for other programming languages and tasks in SE is still unknown and more research is needed.

%% file: texs/07_conclusion.tex
% !TEX root = ../main.tex
\section{Conclusion}\label{sec:conclusion}

In this paper, we explored the generalization ability of popular decoder-only Transformer position encoding schemes that have shown success in Natural Language Processing that can be extended to the encoder-decoder Transformer and code completion task. Specifically, we investigated four different positional encoding schemes, namely Sinusoidal \cite{vaswani2017attention}, xPOS \cite{su2021roformer, sun2022length}, ALiBi \cite{press2022alibi}, and T5 \cite{raffel2020exploring}, which have been proposed as a way to boost this generalization ability and represent the most popular position encoding types, \ie Absolute Positional Encoding and Relative Positional Encoding.

Overall, our results demonstrate that none of the studied positional encoding schemes has the ability to generalize to unseen lengths. 

\eject

While these findings suggest that there are currently no ``shortcuts'' for researchers or developers of tools utilizing these models to efficiently train on short lengths (which are less expensive to process) and generalize to longer lengths, it is interesting to note that training on a mixture of lengths should represent a safe compromise in most of cases. 

Still, the possible drop in performance this may result in should be considered and assessed case by case, depending on the context in which the models must be used, the targeted programming language, and the actual focus on performance.

Moreover, it is worth considering that our conclusions are only based on performance proxies we adopted (\ie EM, ChrF, and RougeL). Different trade-offs come into play if best performance is not the only constraint. For example, while T5 is the best performing positional encoding, it is also the slowest and most memory intensive both for training and inference. Therefore, in performance-critical settings it might be more beneficial to use xPOS, which achieves less performance, but is more efficient. Similar observations can be made for the Sinusoidal and ALiBi schemes for which, however, the cost to pay in terms of performance as compared to T5 is higher.

\eject

Lastly, given our findings and the potential impact that the generalizability of code completions models can have on the software engineering community in terms of training efficiency, we believe future research should explicitly target this problem with research focused outside of different positional encoding schemes and possibly involving additional architectural changes to the Transformer or even proposing completely novel architectures. 

Also, we plan to run similar studies on other code-related tasks, such as code summarization.

\section*{Data Availability}\label{sec:data}
Our datasets, code, models, are available in our reproduction package \cite{replication}. It contains all of our additional results that were excluded due to space constraints. Specifically, additional metrics, \ie BLEU, Levenshtein Distance, Meteor, and Cross Entropy, that follow similar trends as the ones discussed in this paper. 

Additionally, we provide all checkpoints for each of our evaluated models, \ie Sinusoidal, xPOS, ALiBi, and T5, licensed under Apache 2.0. The code repository contains all the code, also released under the same license, for reproducing our results along with documentation on the process required to do so. Lastly, the processed datasets are also released under a creative commons license. We hope that the availability of such code and data can help in fostering more research in this area.

%\section*{Acknowledgments}
% This work is supported by the  grants: AAA and BBB. Any opinions, findings, and conclusions expressed herein are the authors and do not necessarily reflect those of the sponsors.